\documentclass[12pt,reqno]{amsart}

\usepackage{fullpage}
\usepackage{graphicx}
\usepackage{amssymb, amsmath, amsxtra}

\theoremstyle{plain}

\theoremstyle{plain} 

\theoremstyle{plain}

\theoremstyle{plain}

\newtheorem{prop}{Proposition}[section]

\begin{document}
\allowdisplaybreaks[3]

\title{
A nonlinear generalization of\\ 
the Camassa-Holm equation with peakon solutions
}

\author{
Stephen C. Anco$^1$, 
Elena Recio$^{1,2}$, 
Mar\'ia L. Gandarias$^2$, 
Mar\'ia S. Bruz\'on$^2$
\\\lowercase{\scshape{
${}^1$Department of Mathematics and Statistics\\
Brock University\\
St. Catharines, ON L2S3A1, Canada}} \\\\
\lowercase{\scshape{
${}^2$Department of Mathematics\\
Faculty of Sciences, University of C\'adiz\\
Puerto Real, C\'adiz, Spain, 11510}}\\
}

\begin{abstract}
A nonlinearly generalized Camassa-Holm equation,
depending an arbitrary nonlinearity power $p \neq 0$, is considered.
This equation reduces to the Camassa-Holm equation when $p=1$
and shares one of the Hamiltonian structures
of the Camassa-Holm equation.
Two main results are obtained.
A classification of point symmetries is presented 
and a peakon solution is derived,
for all powers $p \neq 0$.
\end{abstract}

\maketitle

\begin{center}
emails: 
sanco@brocku.ca, 
elena.recio@uca.es, 
marialuz.gandarias@uca.es, 
m.bruzon@uca.es
\end{center}

\section{Introduction}

There has been much recent interest 
in nonlinear dispersive equations that model breaking waves. 
One of the first well-studied equations of this kind is 
the Camassa-Holm (CH) equation \cite{CH}
\begin{equation}\label{CH}
u_t -u_{txx} = 3u u_x -2 u_x u_{xx} - u u_{xxx}
\end{equation}
for $u(t,x)$. 
This equation arises from the theory of shallow water waves \cite{CH,CHH}
and provides a model of wave breaking 
for a large class of solutions
in which the wave slope blows up in a finite time
while the wave amplitude remains bounded \cite{C,CE,CE2,CE3}. 
A special class of weak solutions of this equation 
describes peaked solitary waves, known as peakons \cite{ACHM,CH,CHT},
whose wave slope is discontinuous at the wave peak.
More remarkably, 
the CH equation is an integrable system \cite{CH,FS,FF},
possessing a Lax pair, 
a bi-Hamiltonian structure,
and an infinite hierarchy of symmetries and conservation laws. 

Recently in the literature, 
some generalizations of the CH equation \eqref{CH} 
that admit breaking wave solutions
have been studied.
For example,
the $b$-equation \cite{DHH,HH}
\begin{equation}\label{b}
u_t -u_{txx} = - (b+1) u u_x + b u_x u_{xx} + u u_{xxx}
\end{equation}
with parameter $b \neq 0$,
and the 4-parameter family of equations \cite{ASF}
\begin{equation}\label{4par}
u_t -u_{txx} = - a u^p u_x + b u^{p-1} u_x u_{xx} + c u^p u_{xxx}
\end{equation}
with parameters $a,b,c$ (not all zero) and
$p \neq 0$.

In this paper we discuss 
an interesting nonlinear generalization of the CH equation, 
given by
\begin{equation}\label{gCH}
u_t -u_{txx} = 
\tfrac{1}{2} (p+1)(p+2) u^p u_x -\tfrac{1}{2} p(p-1) u^{p-2} u_x{}^3
-2p u^{p-1} u_x u_{xx} -u^p u_{xxx},
\quad
p\neq 0,
\end{equation}
where $p$ is the nonlinearity parameter. 
We call this generalization \eqref{gCH} 
the gCH equation.
It is motivated by 
one of the Hamiltonian structures of the CH equation
and has a close analogy to the relationship between
the generalized Korteweg de Vries (gKdV) equation
$v_t = 
v^p v_x + v_{xxx}$,
with 
$p\neq 0$,
and the ordinary Korteweg de Vries (KdV) equation
$v_t = 
v v_x + v_{xxx}$.
In particular, the generalized KdV equation
reduces to the KdV equation when $p=1$
and shares one of its two Hamiltonian structures.
The same relationship holds between the gCH and CH equations.

In section 2,
we review the Hamiltonian structures of the CH equation
and use one of these structures
to derive the gCH equation \eqref{gCH}.
We also discuss some conservation laws
admitted by the gCH equation.
In section 3, 
we show that 
point symmetries admitted by the gCH equation
consist of translations in $t$ and $x$
and a scaling involving $t$ and $u$.
We use these symmetries
to reduce the gCH equation
to ordinary differential equations
that describe the corresponding group invariant solutions.
In section 4, 
we consider
weak solutions of the gCH equation
and derive a peakon solution
for all nonlinearity powers $p \neq 0$.
We make some concluding remarks in section 5.


\section{Derivation}

The Hamiltonian structures 
of the CH equation \eqref{CH}
are most naturally formulated by 
first introducing the variable
\begin{equation}
m 
= u - u_{xx}.
\end{equation} 
Then the CH equation \eqref{CH}
takes the form of an evolutionary equation
\begin{equation}\label{CHev}
m_t 
= 2 u_x m + u m_x 
= (\tfrac{1}{2}(u^2 - u_x^2) + u m)_x
\end{equation}
for $m(t,x)$,
where $u=\Delta^{-1} m$
is expressed in terms of the operator
\begin{equation}
\Delta 
= 1 - D_x^2.
\end{equation}
This evolutionary equation \eqref{CHev}
has two Hamiltonian structures \cite{CH}
\begin{equation}
m_t
=-\mathcal{H} (\delta H/\delta m)
=-\mathcal{E} (\delta E/\delta m)
\end{equation}
given by the Hamiltonian operators
\begin{equation}\label{HopH}
\mathcal{H}
=m D_x + D_x m
\end{equation}
and
\begin{equation}\label{HopE}
\mathcal{E}
=D_x - D_x^3
=\Delta D_x 
= D_x \Delta,
\end{equation}
where the Hamiltonians are
\begin{equation}
H
={\textstyle\int}_{-\infty}^{+\infty} \tfrac{1}{2} m u\; dx
\end{equation}
and
\begin{equation}
E
={\textstyle\int}_{-\infty}^{+\infty} \tfrac{1}{4} u^2 (u + m)\; dx.
\end{equation} 
Their variational derivatives are computed
by using the relation
\begin{equation}
\dfrac{\delta}{\delta m}
= \Delta^{-1} \dfrac{\delta}{\delta u},
\end{equation} 
with the Hamiltonians expressed
only in terms of $u$ and its $x$ derivatives,
\begin{equation}\label{HamH}
H
={\textstyle\int}_{-\infty}^{+\infty} \tfrac{1}{2} (u^2 + u_x^2)\; dx
\end{equation}
and
\begin{equation}\label{HamE}
E
={\textstyle\int}_{-\infty}^{+\infty} \tfrac{1}{2} u (u^2 + u_x^2)\; dx
\end{equation}
(after dropping total $x$-derivative terms in the integrals).
The Hamiltonian operators \eqref{HopH} and \eqref{HopE}
determine corresponding Poisson brackets defined by
\begin{equation}\label{PoisBr}
\begin{aligned}
\{ \delta F_1/\delta m,\delta F_2/\delta m \}_{\mathcal{D}}  
& = {\textstyle \int}_{-\infty}^{+\infty} 
(\delta F_1/\delta m)\mathcal{D}(\delta F_2/\delta m) dx
\end{aligned}
\end{equation}
in terms of 
$\mathcal{D} = \mathcal{H}$ and 
$\mathcal{D} = \mathcal{E}$, 
where $F_1$ and $F_2$ are arbitrary functionals
in terms of $x, u$ and $x$-derivatives of $u$. 
The bracket \eqref{PoisBr} will be skew
and satisfy the Jacobi identity
if and only if
the operator $\mathcal{D}$ is Hamiltonian \cite{O}. 
One aspect of integrability of the CH equation \eqref{CHev}
is that these two Poisson brackets
given by  $\mathcal{D} = \mathcal{H}$ and $\mathcal{D} = \mathcal{E}$
are compatible in the sense that
any linear combination of them
produces a Poisson bracket.
Correspondingly,
any linear combination of the Hamiltonian operators 
$\mathcal{H}$ and $\mathcal{E}$
is a Hamiltonian operator.

The two Hamiltonians \eqref{HamH} and \eqref{HamE}
of the CH equation \eqref{CHev}
are conserved integrals 
(under suitable asymptotic decay conditions on $u$)
\begin{equation}
\dfrac{d H}{d t}=0,
\quad
\dfrac{d E}{d t}=0
\end{equation}
due to the antisymmetry of the Poisson brackets \eqref{PoisBr}.
Since the CH equation \eqref{CHev} itself
is in the form of a conservation law,
the integral
\begin{equation}\label{HamP}
P
={\textstyle\int}_{-\infty}^{+\infty} m\; dx
\end{equation}
is also conserved,
\begin{equation}\label{clP}
\dfrac{d P}{d t}=0.
\end{equation}
There is a third Hamiltonian structure
of the CH equation \eqref{CHev}
for which this integral \eqref{HamP}
is the Hamiltonian \cite{HW}.

It is useful to observe
that the second Hamiltonian structure 
given by the operator
\eqref{HopE}
can be equivalently expressed
in a strictly local variational form
in terms of $u$ through the identity
\begin{equation}
\mathcal{E} (\delta F / \delta m) 
= D_x (\delta F / \delta u)
\end{equation}
(which holds for any functional $F$).
This formulation gives
\begin{equation}\label{CHvf}
m_t=-D_x(\delta E / \delta u)
\end{equation}
where $E$ is the Hamiltonian \eqref{HamE}.

A natural nonlinear generalization
of the variational formulation \eqref{CHvf} 
consists of simply replacing the Hamiltonian \eqref{HamE} by
\begin{equation}\label{HamEp}
E_{(p)}
={\textstyle\int}_{-\infty}^{+\infty}  \tfrac{1}{2} u^p (u^2 + u_x^2)\; dx,
\quad
p \neq 0,
\end{equation} 
which yields the Hamiltonian evolutionary equation
\begin{equation}\label{gCHvf}
m_t =
- D_x (\delta E_{(p)} / \delta u) =
- \mathcal{E} (\delta E_{(p)} / \delta m),
\end{equation}
where $p$ is an arbitrary nonlinearity power.
In this formulation, the Hamiltonian \eqref{HamEp}
can be equivalently expressed
in terms of $u$ and $m$,
as given by
\begin{equation}
E_{(p)}
={\textstyle\int}_{-\infty}^{+\infty} \tfrac{1}{2} (p+1)^{-1} u^{p+1} (p u + m)\; dx.
\end{equation}
(after dropping a total $x$-derivative term).
The gCH equation \eqref{gCHvf}
reduces to the CH equation \eqref{CHev} when $p=1$. 
For $p \neq 1$, 
the gCH equation \eqref{gCHvf}
is a nonlinear variant of the CH equation \eqref{CHev},
analogous to how the gKdV equation
nonlinearly generalizes the KdV equation.

Like the CH equation, 
the gCH equation \eqref{gCH}
is in the form of a conservation law
\begin{equation}
u_t - u_{txx} =
(\tfrac{1}{2} p u^{p-1} (u^2 - u_x^2)+u^p (u - u_{xx}))_x.
\end{equation}
Thus the integral \eqref{clP} is conserved
(under suitable asymptotic decay conditions on $u$).
Another conserved integral is provided by the Hamiltonian \eqref{HamEp}.
This integral gives rise to the conservation law
\begin{equation}
D_t T + D_x X = 0,
\end{equation}
with

\begin{equation}
\begin{aligned}
&
T =
\tfrac{1}{2} u^p (u^2 + u_x^2),
\\
&
X = 
- u^p u_t u_x + \tfrac{1}{2} (\delta E / \delta m)^2
- \tfrac{1}{2} (D_x(\delta E / \delta m))^2,
\end{aligned}
\end{equation}
where
\begin{equation}
\delta E / \delta m =
\Delta^{-1} (\delta E / \delta u) =
\Delta^{-1} (\tfrac{1}{2} p u^{p-1} (u^2 - u_x^2)+u^p (u - u_{xx})).
\end{equation}
We will present 
a complete classification of conservation laws 
of the gCH equation \eqref{gCH} elsewhere.


\section{Symmetry Analysis}

We will now consider the gCH equation \eqref{gCH} 
written in the form of a system
\begin{equation}\label{sys1}
\begin{aligned}
& m 
= u - u_{xx},
\\ 
& m_t
= 2 p u^{p-1} u_x m + u^p m_x + \tfrac{1}{2} p (p-1) u^{p-2} (u^2 - u_x^2) u_x
\end{aligned}
\end{equation}
for $u(t,x)$, $m(t,x)$.

A point symmetry \cite{BA,O} of system \eqref{sys1} is
a one-parameter Lie group of transformations on $(t,x,u,m)$ 
generated by a vector field of the form 
\begin{equation}\label{vect2}
{\bf X}=\tau(t,x,u,m)\partial_t
+ \xi(t,x,u,m)\partial_x 
+ \eta(t,x,u,m)\partial_u 
+ \phi(t,x,u,m)\partial_m
\end{equation} 
which is required to leave invariant 
the solution space of system \eqref{sys1}. 
The condition of invariance is given by
applying the prolongation of ${\bf X}$
to each equation in the system \eqref{sys1}.
After prolongation, the resulting equations
split with respect to the $t$ and $x$ derivatives
of $u$ and $m$, 
yielding an overdetermined, linear system of $47$ equations 
for $\tau(t,x,u,m)$, $\xi(t,x,u,m)$, $\eta(t,x,u,m)$, $\phi(t,x,u,m)$,
together with the parameter $p$. 
We derive and solve this linear system
by using the Maple package GeM \cite{Ch}. 

\begin{prop}
The infinitesimal point symmetries
admitted by the gCH system \eqref{sys1} for $p \neq 0$
are generated by
\begin{align}
\label{symx}
& {\bf X}_1 =  \partial_x,
\quad
\text{translation in x,}
\\\label{symt}
& {\bf X}_2 =  \partial_t, 
\quad 
\text{translation in t,}
\\\label{symscal}
& {\bf X}_3 = m \partial_m + u \partial_u - p t \partial_t,
\quad 
\text{scaling,}
\end{align}
(there are no extra symmetries admitted only for special values of $p \neq 0$).
All of these symmetries \eqref{symx}--\eqref{symscal} 
project to point symmetries of the gCH equation \eqref{gCH}.
\end{prop}

Each admitted point symmetry \eqref{vect2} can be used 
to reduce the gCH system (\ref{sys1})
to a system of ordinary differential equations  (ODEs)
whose solutions correspond to invariant solutions 
$(u(t,x),m(t,x))$ 
of system (\ref{sys1})
under the point symmetry. 
These invariant solutions are naturally expressed
in terms of similarity variables
\begin{equation}
(z(t,x),U(t,x,u,m),M(t,x,u,m)),
\end{equation} 
which are found by solving the invariance conditions
\begin{equation}
\begin{aligned}\label{ic}
& \eta(t,x,u,m)-\tau(t,x,u,m)u_t-\xi(t,x,u,m)u_x = 0,
\\ 
& \phi(t,x,u,m)-\tau(t,x,u,m)m_t-\xi(t,x,u,m)m_x = 0,
\end{aligned}
\end{equation}
with $U_u \neq 0$ and $M_m \neq 0$.


\subsection{Reduction under translations}

Reductions under the separate translation symmetries 
\eqref{symx} and \eqref{symt} are not interesting. 
Instead we consider the combined space-time translation symmetry
\begin{equation}
{\bf X}=\partial_t - c \partial_x,
\end{equation}
where $c$ is a non-zero constant. For this symmetry
the invariance conditions \eqref{ic} are given by
\begin{equation}
u_t-cu_x=0,
\quad
m_t-cm_x=0,
\end{equation}
which yields the similarity variables
\begin{equation}
z = x + c t,
\quad
U=u,
\quad
M=m.
\end{equation}
The resulting form for invariant solutions
of the gCH system \eqref{sys1}
is a travelling wave given by
\begin{equation}
u = U(z),
\quad
m = M(z),
\end{equation}
satisfying the ODE system
\begin{equation}\label{sysODEtw}
\begin{aligned}
M = & \ U-U'',
\\ 
c M'= & \
\tfrac{1}{2} p (U^{p-1} (U^2-U'^2))'+(M U^p)'.
\end{aligned}
\end{equation}
This system \eqref{sysODEtw} 
is equivalent to the nonlinear third order ODE
\begin{equation}\label{ODEtw0}
0 = (\tfrac{1}{2}p U^{p-1}(U^2-U'^2)+(-c + U^p)(U-U'') )'.
\end{equation}

We look for solutions $U(z)$
that describe solitary waves, 
as characterized by the asymptotic boundary conditions
\begin{equation}\label{acODEtw}
U, U', U'' \rightarrow 0 
\quad
\text{for}
\quad
\lvert z \rvert \rightarrow \infty.
\end{equation}
The ODE \eqref{ODEtw0} has the obvious first integral
\begin{equation}
\tfrac{1}{2}p U^{p-1}(U^2-U'^2)+ (-c + U^p)(U-U'') =
\alpha = \text{const.}
\end{equation}
Imposing the asymptotic conditions \eqref{acODEtw}, 
we have
\begin{equation}
\alpha = 0.
\end{equation}
The resulting second order ODE is
\begin{equation}\label{ODEtw}
\tfrac{1}{2}p U^{p-1}(U^2-U'^2)+(-c + U^p)(U-U'') =
0.
\end{equation}
This ODE \eqref{ODEtw} has an integrating factor $U'$,
which yields the first integral
\begin{equation}
(-c + U^p)(U^2-U'^2) =
\beta = \text{const.}
\end{equation}
Again from imposing the asymptotic conditions \eqref{acODEtw},
we get
\begin{equation}
\beta = 0.
\end{equation}
Thus, we obtain
\begin{equation}\label{peakeqn}
(-c + U^p)(U^2-U'^2) =
0,
\end{equation}
which implies that
$U = \pm U'$ whenever $U^p \neq c$.
Clearly, no smooth function $U(z)$
can satisfy both 
$U' = \pm U$ and $U^p \neq c$
such that $U \rightarrow 0$ 
as $ \lvert z \rvert \rightarrow \infty $.
Thus there do not exist 
any smooth, asymptotically decaying solutions 
of the travelling wave ODE \eqref{ODEtw0}.

This analysis suggests that
we look for weak solutions, 
called peakons,
\begin{equation}\label{peak}
U(z)=c^{1/p} \exp(-\lvert z \rvert),
\end{equation}
obeying the asymptotic conditions \eqref{acODEtw}.
In the next section,
we will show that this expression \eqref{peak}
does satisfy the weak form
of the travelling wave ODE \eqref{ODEtw0}.

\begin{prop}
The gCH equation \eqref{gCH} admits
peaked travelling waves (which are weak solutions)
\begin{equation}\label{peak2}
u=c^{1/p} \exp(-\lvert x+ct \rvert),
\quad
p \neq 0,
\end{equation}
where $c$ is an arbitrary constant.
\end{prop}

When $p=1$,
this result coincides with 
the well-known peakon solution
$u=c  \exp(-\lvert x+ct \rvert)$
of the CH equation \eqref{CH}.
When $p \neq 1$ is an odd integer,
or more generally
when $p$ is rational with an odd denominator,
then the peakon solution \eqref{peak2}
holds with no restriction
on the sign of $c$.
In all other cases, $c$ must be positive.


\subsection{Reduction under scaling}

For the scaling symmetry \eqref{symscal}, 
the invariance conditions \eqref{ic} are given by
\begin{equation}
u+ptu_t=0,
\quad
m+ptm_t=0,
\end{equation}
which yields the similarity variables
\begin{equation}
z=x,
\quad
U=t^{1/p}u,
\quad
M=t^{1/p}m.
\end{equation}
The resulting form for scaling-invariant solutions
of the gCH system \eqref{sys1}
is given by
\begin{equation}
u = t^{-1/p} U(z),
\quad
m = t^{-1/p} M(z),
\end{equation}
satisfying the ODE system
\begin{equation}\label{sysODEsim}
\begin{aligned}
M = & \ U-U'',
\\ 
- p^{-1} M= & \
(-\tfrac{1}{2} p U^{p-1} (U^2-U'^2)-U^p M)'.
\end{aligned}
\end{equation}
This system \eqref{sysODEsim} is equivalent to
the nonlinear third order ODE
\begin{equation}\label{ODEsim}
0 = p^{-1}(U-U'') - (\tfrac{1}{2}pU^{p-1}(U^2-U'^2)+U^p(U-U''))'
\end{equation}
for $U(z)$.
Solutions of this ODE \eqref{ODEsim}
yield similarity solutions
\begin{equation}\label{ODEsimsol}
u=t^{-1/p}U(x)
\end{equation}
of the gCH equation \eqref{gCH}.

When $p>0$, 
these similarity solutions \eqref{ODEsimsol} will exhibit 
decay $u \rightarrow 0$ as $t \rightarrow \infty$ 
and a blow-up $u \rightarrow \infty$ at $t=0$.
By applying a time translation,
we will get a solution
\begin{equation}
u=(t-t_0)^{-1/p} U(x),
\quad
t_0 = \text{const.}
\end{equation}
which still decays to $0$ for large $t$
but has no blow-up for $t \geq 0$ when $t_0 < 0$. 
We would like to find solutions 
that have spatial decay $u \rightarrow 0$ for large $x$.
Correspondingly,
we want solutions $U(z)$ 
of the similarity ODE \eqref{ODEsim}
that satisfy the asymptotic boundary conditions \eqref{acODEtw}.

The similarity ODE \eqref{ODEsim}
can be shown to admit no point symmetries
other than translations in $z$,
and no contact symmetries,
as well as no integrating factors
at most linear in $U''$.
The $z$-translation symmetry yields
only a trivial invariant solution,
$U(z)=0$.
Thus, standard integration methods
fail to yield any non-trivial solutions of ODE \eqref{ODEsim}.
By inspection,
however, we see that 
$U(z)=a \exp(\pm z)$
is an exact solution of ODE \eqref{ODEsim},
where $a$ is an arbitrary constant.
These exponential solutions fail
to satisfy the asymptotic conditions \eqref{acODEtw},
but this might suggest looking for a peakon solution
$U(z)=a \exp (-\lvert z \rvert)$.
However,
we will show in the next section
that the weak form of the similarity ODE \eqref{ODEsim}
does not admit such solutions.


\subsection{Other reductions.}

If two symmetry generators \eqref{vect2}
are related by conjugation
with respect to some subgroup
in the full group of point symmetry
generated by $X_1, X_2, X_3$,
then the action of this symmetry subgroup on solutions
$(u(t,x), m(t,x))$
will map the group-invariant solutions
determined by the two symmetry generators 
into each other.
Consequently,
for the purpose of finding 
all group-invariant solutions,
it is sufficient to work 
with any maximal set of symmetry generators
that are conjugacy inequivalent.
From the point symmetry algebra
$[X_1,X_2]=[X_1,X_3]=0$, $[X_2,X_3]=-pX_2$,
the equivalence classes under conjugation
consist of $X_1+aX_2,X_3+bX_1$,
where $a,b$ are arbitrary parameters.
We consider the reductions with 
$a \neq 0$ and $b \neq 0$ elsewhere.


\section{Peakon Solutions}

To show that the peakon \eqref{peak}
is a weak solution
of the travelling wave ODE \eqref{ODEtw0},
we start from an equivalent integral formulation
obtained by multiplying the ODE with a test function $\psi$
(which is smooth and has compact support)
and integrating over
$-\infty < z < \infty $,
leaving at most first derivatives of $\psi$ in the integral.
This yields,
after integration by parts,
\begin{equation}\label{inteqn}
0 =\int_{-\infty}^{+\infty} \big(-c (\phi U + \phi'U') +
\tfrac{1}{2} p \phi U^{p-1}(U^2-U'^2) + 
\phi U^{p+1} + U'(\phi U^p)'\big)  \; dz,
\end{equation}
where $\phi = \psi'$ is also a test function.
Weak solutions of ODE \eqref{ODEtw} 
are functions $U(z)$ that belong 
to the Sobolev space $W_{\text{loc}}^{1,p+1}(\mathbb{R})$
and that satisfy the integral equation \eqref{inteqn}
for all smooth test functions $\phi(z)$
with compact support.

Now we substitute a peakon expression 
\begin{equation}\label{peakODE}
U=ae^{-\lvert z \rvert}, 
\quad
U'=-a {\rm sgn} (z) e^{-\lvert z \rvert}
\end{equation}
into equation \eqref{inteqn}
and split up the integral into the intervals 
$- \infty < z \leq 0$ and $0 \leq z < +\infty$. 
From the first term in equation \eqref{inteqn},
after integrating by parts,
we obtain
\begin{equation}\label{inteqn1}
- \int_{-\infty}^{0} c(\phi U + \phi' U') \; dz 
- \int_{0}^{+\infty} c(\phi U + \phi' U') \; dz =
- 2ac \phi(0).
\end{equation}
The second term in equation \eqref{inteqn}
is $0$ since the peakon expression \eqref{peakODE} 
satisfies $U^2 = U'^2$.
Expanding the fourth term in equation \eqref{inteqn},
we have
\begin{equation}\label{inteqn3}
\int_{-\infty}^{+\infty} (\phi' U'U^p + p \phi U^{p-1}U'^2) \; dz.
\end{equation}
The first term in the integral \eqref{inteqn3} yields,
after integration by parts,
\begin{equation}\label{inteqn32}
\int_{-\infty}^{0}\phi' U'U^p  \; dz+
\int_{0}^{+\infty}\phi' U'U^p  \; dz=
2 a^{p+1}\phi(0)-a^{p+1}(p+1)\int_{-\infty}^{+\infty} \phi e^{-(p+1)\lvert z \rvert} \; dz.
\end{equation}
The second term in the integral \eqref{inteqn3} yields
\begin{equation}\label{inteqn33}
\int_{-\infty}^{+\infty} p \phi U^{p-1}U'^2 \; dz=
p a^{p+1} \int_{-\infty}^{+\infty} \phi e^{-(p+1)\lvert z \rvert} \; dz.
\end{equation}
Similarly,
the third term in the integral \eqref{inteqn} yields
\begin{equation}\label{inteqn31}
\int_{-\infty}^{+\infty} \phi U^{p+1} \; dz= a^{p+1}
\int_{-\infty}^{+\infty} \phi e^{-(p+1)\lvert z \rvert} \; dz.
\end{equation}
Combining the terms \eqref{inteqn1}--\eqref{inteqn31},
we get
\begin{equation}
2 (- ac + a^{p+1})\phi(0) = 0.
\end{equation}
Since $\phi(0)$ is arbitrary,
this implies
\begin{equation}
a = c^{1/p},
\end{equation}
which establishes that 
the peakon \eqref{peak} satisfies equation \eqref{inteqn}.

The integral formulation 
of the similarity ODE \eqref{ODEsim}
is given by
\begin{equation}\label{ODEsim_ieq}
0 =\int_{-\infty}^{+\infty} \phi \big(p^{-1}(U-U'') - (\tfrac{1}{2}pU^{p-1}(U^2-U'^2)+U^p(U-U''))'\big)  \; dz,
\end{equation}
where $\phi(z)$ is a test function.
After integrating by parts,
we obtain
\begin{equation}\label{ODEsim_ieq2}
0 =\int_{-\infty}^{+\infty} \big( p^{-1}(\phi U +
\phi'U') + 
\tfrac{1}{2} p \phi' U^{p-1}(U^2-U'^2) +
\phi' U^{p+1} +
(\phi' U^p)'U'
\big)  \; dz.
\end{equation}
We substitute a peakon expression \eqref{peakODE} 
into equation \eqref{ODEsim_ieq2}
and split up the integral into the intervals 
$- \infty < z \leq 0$ and $0 \leq z < +\infty$. 
From the first term in equation \eqref{ODEsim_ieq2},
after integrating by parts, we obtain
\begin{equation}\label{ieqn21}
\int_{-\infty}^{0} p^{-1}(\phi U + \phi'U') \; dz +
\int_{0}^{+\infty} p^{-1}(\phi U + \phi'U') \; dz =
2 ap^{-1} \phi(0).
\end{equation}
The second term in equation \eqref{ODEsim_ieq2} is $0$
since the peakon expression \eqref{peakODE}
satisfies $U^2 = U'^2$.
The third and fourth terms in equation \eqref{ODEsim_ieq2}
yield, after integration by parts,
\begin{equation}\label{ieqn22}
\int_{-\infty}^{0} \big( \phi' U^{p+1} + (\phi' U^p)'U' \big) \; dz +
\int_{0}^{+\infty} \big( \phi' U^{p+1} + (\phi' U^p)'U' \big) \; dz =
2 a^{p+1} \phi'(0).
\end{equation}
Combining the terms \eqref{ieqn21}--\eqref{ieqn22}, we get
\begin{equation}
2ap^{-1} \phi(0) + 2 a^{p+1} \phi'(0) = 0.
\end{equation}
Since $\phi$ is arbitrary, this implies
\begin{equation}
a = 0.
\end{equation}
Thus, the weak form of the similarity ODE \eqref{ODEsim}
does not admit peakon solutions \eqref{peak}.


\section{Remarks}

We have introduced 
a nonlinearly generalized CH equation \eqref{gCH},
depending on an arbitrary nonlinearity power $p \neq 0$.
This equation reduces to the CH equation when $p=1$
and shares one of the Hamiltonian structures of CH equation \eqref{CH}.
For all $p \neq 0$, 
it admits a peakon solution \eqref{peak2}.

The gCH equation is worth further study
to understand how its nonlinearity affects
properties of its solutions 
compared to the CH equation.
In particular,
the CH equation is
an integrable system,
admits multi-peakon weak solutions,
and exhibits wave-breaking
for a large class of classical solutions.
Is the gCH equation integrable
for some nonlinearity power $p \neq 1$?
Does it admit multi-peakon solutions
for nonlinearity powers $p \neq 1$?
Is it well-posed for all $p \neq 1$?
Does it exhibit the same wave-breaking behavior
for all $p \neq 1$?
Is there a critical power $p$
for which a different kind of blow-up occurs
(other than wave breaking)?

\end{document}